\documentstyle[12pt,epsfig]{article}
\begin{document}
\setlength{\headheight}{0in}
\setlength{\headsep}{0in}
\setlength{\topskip}{1ex}
\setlength{\textheight}{8.5in}
\setlength{\topmargin}{0.5cm}
\setlength{\baselineskip}{0.24in}
%%%%%%%%%%%%%%%
\catcode`@=11
% Redefine caption to put text and formulas in smaller font
\long\def\@caption#1[#2]#3{\par\addcontentsline{\csname
  ext@#1\endcsname}{#1}{\protect\numberline{\csname
  the#1\endcsname}{\ignorespaces #2}}\begingroup
    \small
    \@parboxrestore
    \@makecaption{\csname fnum@#1\endcsname}{\ignorespaces #3}\par
  \endgroup}
\catcode`@=12
%%%%%%%%%%%%%%%%%%%%%%%%%%%%%%%%%%%%%%%%%%%%%%%%%%%%%%%%%%%%
\def\slashchar#1{\setbox0=\hbox{$#1$}           % set a box for #1
   \dimen0=\wd0                                 % and get its size 
   \setbox1=\hbox{/} \dimen1=\wd1               % get size of /
   \ifdim\dimen0>\dimen1                        % #1 is bigger 
      \rlap{\hbox to \dimen0{\hfil/\hfil}}      % so center / in box 
      #1                                        % and print #1
   \else                                        % / is bigger 
      \rlap{\hbox to \dimen1{\hfil$#1$\hfil}}   % so center #1
      /                                         % and print /
   \fi}                                         %
\newcommand{\newc}{\newcommand}
\def\be{\begin{equation}}
\def\ee{\end{equation}}
\def\bea{\begin{eqnarray}}
\def\eea{\end{eqnarray}}
\def\simlt{\stackrel{<}{{}_\sim}}
\def\simgt{\stackrel{>}{{}_\sim}}
\begin{titlepage}
\vskip 2cm
\begin{center}

{\Large\bf 
The Higgs masses  and  explicit CP violation in the gluino-axion model}

\vskip 1cm

{\large
M\"{u}ge Boz \\}

\vskip 0.5cm
{\setlength{\baselineskip}{0.18in}
Physics Department, Hacettepe University, 06532, Beytepe, Ankara\\}

\end{center}
\vskip .5cm
\begin{abstract}
In this work,  we  adress the phenomenological consequences 
of explicit CP violation on direct Higgs-boson searches at high energy 
colliders. Having a  restricted parameter space, we concentrate on the recently proposed
gluino-axion model, and investigate the CP violation capability 
of the model subject to the recent experimental data.  
It is shown that the Higgs masses as well as their CP compositions are quite sensitive
to the supersymmetric CP phases. The lightest Higgs is found to be
nearly CP even to a good approximation whilst  the remaining two heavy
scalars do not have definite CP parities.
\end{abstract}
\end{titlepage}
%%%%%%%%%%%%%%%%%%%%%%%%%%%%%%%%%%%%%%%%%%%%%%%%%%%%%%%%%%
\setcounter{footnote}{0}
\setcounter{page}{1}
\setcounter{section}{0}
\setcounter{subsection}{0}
\setcounter{subsubsection}{0}
%%%%%%%%%%%%%%%%%%%%%%%%%%%%%%%%%%%%%%%%%%%%%%%%%%%%%%%%%%%%%%%%%%%%%%%

\section{Introduction}
Presently, the phenomenon of CP non-conservation is one of the key problems from
both theoretical and experimental points of views. The observed CP violation in 
neutral kaon system \cite{1} as well as the electric dipole moment (EDM) of the 
neutron \cite{2} severely constrain the sources and strength of CP violation
in the underlying model. In the standard model (SM) both strong and electroweak 
interactions violate the CP invariance. It is a well--known fact that the
$\theta$ vacuum \cite{3} violates the CP invariance, and results in 
a neutron EDM exceeding the present bounds by nine orders of magnitude \cite{4}.
This is the source of the strong CP problem -- a CP hierarchy and naturalness 
problem.
 
In the supersymmetric (SUSY) extensions of the standard electroweak theory (SM) 
this hierarchy problem still persists. Moreover, there appear novel sources of 
CP violation coming from the soft supersymmetry breaking mass terms. Though 
the phases of the soft terms have been shown to relax to CP--conserving points
in the minimal model (MSSM) \cite{5}, this is not necessarily true in the 
non--minimal model (NMSSM) \cite{6} containing a singlet. These
soft terms contribute to known CP--violating observables \cite{7} (EDM's and neutral 
meson mixings); however, they also induce CP violation in the Higgs
sector\cite{8,9,10,11}.

In addition to these CP hierarchy problems, in minimal SUSY 
model there is another hierarchy problem concerning the Higgsino Dirac 
mass parameter ($\mu$), that is, this mass parameter follows from the 
superpotential of the model and there is no telling of at what 
scale (ranging from $M_W$ to $M_{Pl}$) it is stabilized.

In a recently proposed  model so called as gluino-axion\cite{12,13}
the two hierarchy problems, i.e, the strong CP   
problem and the $\mu$ problem are  solved in the context of supersymmetry
with a new kind of axion \cite{14,15} which couples to the gluino
rather than to quarks. In this model
the invariance of the supersymmetric Lagrangian and all supersymmetry
breaking terms  under $U(1)_R$ is guaranteed  by  promoting the ordinary   
$\mu$ parameter to a composite operator involving the gauge singlet $\hat{S}$
with unit $R$ charge. When the scalar component of the singlet develops vacuum
expectation value (VEV) around the Peccei--Quinn scale $\sim 10^{11}~{GeV}$  an
effective $\mu$ parameter $\mu\sim \mbox{a TeV}$ is induced. Besides, the
low energy theory is identical to minimal SUSY  model  with all 
sources of soft supersymmetric phases. Due to all these abilities of the
model of Ref.~12 in solving the hierarchy
problems, in the analysis below we will adopt its
parameter space. 

In this work, we adress the consequences of explicit CP violation 
on the the radiatively corrected Higgs masses and mixings in the framework
of the gluino-axion model. It will
be seen that the supersymmetric phases significantly affect the Higgs masses
and mixings thereby giving new regions in the parameter space (otherwise
excluded) meeting the recent LEP constraints \cite{16}.
As a result of  the standart model Higgs boson
searches at LEP, the lower bound on the lightest Higgs mass
is $115~{\rm GeV}$ (and correspondingly $\tan\beta \simgt 3.5 $)\cite{16}.
On the other hand, theoretically  the lightest Higgs boson mass can not exceed
$130~{\rm GeV}$ for large $\tan\beta$ \cite{17}. Therefore, from the searches at LEP2,
the lower limit on mass of the SM Higgs boson excludes the
substantial part of the MSSM parameter space particularly at
small $\tan\beta$ ($\tan\beta \simlt 3.5$)\cite{16}.

It is a well-known fact that, CP is conserved  in  the Higgs sector
of the minimal supersymmetric model (MSSM) at the tree level. 
On the other hand, the radiative corrections to the masses of the Higgs bosons, 
dominated by top-quark and top-squark loops,
have been found to modify significantly the tree level bound \cite{18,19}. 
The CP conserving Higgs sector has been
analyzed  by several authors and  the radiative corrections
which make very important contributions to the Higgs masses 
have been computed by using different approximations such as  diagrammatic \cite{18,20}
and effective potential methods \cite{19}. 
More complete treatment  of these results include the  
complete one-loop on-shell renormalization  
\cite{21}, the renormalization group (RG) improvement for resumming the leading logarithms\cite{22},
the iteration of the RG  equations to two--loops
with the use of the effective potential techniques\cite{23}
and  the two loop on-shell renormalization \cite{17,24}.

On the other hand, as is indicated  in Ref.~9 that
the explicit CP violation in the matrices of third generation squarks can induce CP violation through
loop corrections. In the recent literature, the radiatively induced CP violation effects has been
studied without \cite{9,10,11,25} or with \cite{26,27} RG
improvement. In between these works, the diagrammatic computation of the
scalar-pseudoscalar transitions was the the scope of Ref.~9, where  
the implications of  the presence of CP phases in the soft
SUSY breaking sector allowing to the mixing of CP even and CP odd states
were discussed. More recently, the  mass matrix of the neutral Higgs bosons 
of the MSSM has been calculated with the effective potential method
in  Refs.~10 and 11 from different perspectives.   
The  detailed analysis of underlying dynamics under study 
is performed  in case of the small splittings between 
squark mass eigenstates in Ref.~10;
whereas  bottom-sbottom contributions  are not taken
into account in Ref.~11.
Additional contributions from the chargino, W and  the charged Higgs exchange loops 
were computed in Ref.~25.
In Ref.~26, one-loop corrections to the mass matrix of the neutral Higgs bosons in the
MSSM were calculated by using the effective potential method for an 
arbitrary splitting between squark masses,  including the electroweak and
gauge couplings and the leading two loop corrections.
Although, the earlier works \cite{10,11} on the Higgs spectrum 
were based on some approximations, the results are in agreement with the ones
presented Ref.~26 in the appropriate limit. 
More complete treatment of the effective Higgs potential in the MSSM
including the two-loop leading logarithms induced by top-bottom
Yukawa couplings as well as those associated with QCD corrections by
means of RG methods were performed in Ref.~27 in which 
the leading logarithms generated by one-loop gaugino and higgsino
quantum effects are also taken into consideration.

We would like to point out that, it is the main purpose of this work to
investigate the CP violation capability of the gluino-axion model. Therefore,
we limit our analysis to the effective potential with no RG
improvement. This accuracy has proven sufficient in obtaining the observable
effects of explicit CP violation on the Higgs masses and mixings
\cite{11}. In the following, we compute the  radiatively corrected Higgs
masses and mixings, taking into account the  CP violation effects.
We will base our calculations to those of Ref.~11,
by modifying the parameters appropriately in connection with the gluino- axion model.
The  main difference with the previous work \cite{8,11}
springs from the fact that the parameters chosen are specific to the gluino-axion model, namely 
all  the soft mass parameters in this theory are fixed in terms of the $\mu$ parameter.

The organization of this work is as follows: In Sec. 2,  starting from the
Higgs sector structure of the gluino-axion model, we compute the
$(3\times3)$ dimensional mass matrix of the Higgs scalars in which
all the elements  are expressed in terms of the parameters of  the model
under concern.
In Sec. 3,  we make the numerical
analysis for evaluating the masses of the  Higgs bosons and analyzing
the relative strengts of CP-violating and CP-conserving mixings.
In Sec. 4, we conclude the work.

\section{Higgs Sector in the Gluino-Axion Model}

In this section, our starting point will be the description of the  basic 
low-energy structure of the gluino-axion model which contains the sources of
explicit CP violation. 
The  gluino-axion model, is defined by the superpotential 
\begin{eqnarray}
\widehat{W}=\mu(\widehat{S})\widehat{ H}_{u}.\widehat{H}_{d}+m_{s}^2\mu(\widehat{S})
+h_{u}\widehat{Q}.\widehat{H}_{u}\widehat{u}_{c}
+h_{d}\widehat{Q}.\widehat{H}_{d}\widehat{d}_{c}
+h_{e}\widehat{L}.\widehat{H}_{d}\widehat{e}_{c}
\end{eqnarray}
where $\widehat{Q},\,\,\widehat{u}_{c},\,\,\widehat{d}_{c},\,\,
\widehat{L},\,\,\widehat{e}_{c}$
are the quark, lepton 
and $\widehat{H}_{u},\,\,\widehat{H}_{d}$
are the  Higgs  superfields respectively. The model,
replaces $\mu$ with the  composite operator containing the singlet
composite superfield  $\widehat{S}$ of  R=+1 so that the resulting supersymmetric
Lagrangian and  all supersymmetry breaking terms are  invariant  under $U(1)_{R}$ \cite{12}.
The pure singlet contribution $m_{s}^2\mu(\widehat{S})$
in $\widehat{W}$ is allowed by the symmetries of the model. 

The soft terms of the 
low energy Lagrangian in the gluino-axion model are identical to those in the 
general MSSM \footnote{Here and in what follows we will
neglect the effects of axion, axino, and saxino as their
couplings are severely suppressed \cite{12}.}
\begin{eqnarray}
\label{softMSSM}
{\cal{L}}^{soft}_{MSSM}&=& \tilde{Q}^{\dagger}M_{Q}^{2} \tilde{Q} + 
\tilde{u^c}^{\dagger} M_{u^c}^{2} \tilde{u^c}+\tilde{d^c}^{\dagger}
M_{d^c}^{2}\tilde{d^c}+\tilde{L}^{\dagger} M_{L}^{2} \tilde{L}+
\tilde{e^c}^{\dagger} M_{e^c}^{2} \tilde{e^c}\nonumber\\&+&
\Big\{ A_{u} \tilde{Q}\cdot {H}_{u}~\tilde{u^c}+A_{d} \tilde{Q}\cdot 
{H}_{d} ~\tilde{d^c} + A_{e} \tilde{L}\cdot {H}_{d} ~\tilde{e^c}\big] + h. c.
\Big\}\nonumber \\ &+& M_{H_u}^{2} |H_u|^{2}+M_{H_d}^{2} |H_d|^{2}+
\left(\mu\ B H_{u}\cdot H_{d} + h. c. \right)\nonumber\\
&+&\Big\{M_{3} \tilde{\lambda}^{a}_{3}\tilde{\lambda}^{a}_{3} +
M_2 \tilde{\lambda}^{i}_{2}\tilde{\lambda}^{i}_{2}+ M_1 \tilde{\lambda}_{1}
\tilde{\lambda}_{1}+ h. c. \Big\},
\end{eqnarray}
except for the fact that the soft masses are all expressed in terms of
the $\mu$ parameter through appropriate flavour matrices. The flavour 
matrices form the sources of CP violation and intergenerational mixings in 
the squark sector. The phases of the trilinear couplings ($A_{u,d,e}$), 
the gaugino masses ($M_{3,2,1}$), and the effective $\mu$--parameter 
\begin{eqnarray}
\mu\equiv v_s^2/M_{Pl}\times e^{-i \theta_{QCD}/3}\sim \mbox{a TeV}\times e^{-i \theta_{QCD}/3}
\end{eqnarray}
are the only phases which can generate CP violation observables. In this formula
for the $\mu$ parameter $v_s\sim 10^{11}~{\rm GeV}$ is the Peccei--Quinn scale, and  $\theta_{QCD}$ is 
the effective QCD vacuum angle. 
One notes that the vacuum expectation value of the singlet serves for two important
purposes for the model under concern: Its magnitude determines the scale of supersymmetry breaking and
its phase solves the strong CP--problem.

In the following, we shall calculate the one-loop corrections to the Higgs 
masses and mixings. In doing this, we will modify the parameters in
connection with the gluino-axion model. As in the CP-conserving case \cite{18,19}, 
among the particles contributing to the one-loop radiative corrections, 
the dominant ones come from the top quark and top squark loops provided that 
$\tan\beta\simlt 50$ (in which case the bottom Yukawa coupling is too small to give 
significant contributions). The Yukawa interactions due to scalar-bottom
quarks can be significant only for very large $\tan\beta$ values. On the
other hand,  the contributions of gauginos
and Higgsinos are  already negligible  since they  couple via  weak coupling.
In our analysis we restrict ourselves for the case $\tan\beta\simlt 50$,
so that the dominant terms will be given by the top quark and top squark
loops, to a good approximation.
Therefore,  we will not need the full flavour structures in 
Ref.~12, instead we will need to specify only the top squark sector: 

$(i)$ The top squark soft masses:
\begin{eqnarray}
M_{\tilde{Q}}^{2}=k_{Q}^{2}\ |\mu|^{2}~,\ \ \ M_{\tilde{u}}^{2}= k_{u}^{2}\ |\mu |^{2}~,\ \ \ M_{\tilde{d}}^{2}=
k_{d}^{2}\ |\mu |^{2}
\end{eqnarray} 
where $k_{Q,u,d}$ are real parameters. 

$(ii)$ The top squark trilinear coupling
\begin{eqnarray}
\label{at}
A_{t}=\mu^{*}\ k_t,
\end{eqnarray}
where $k_t$ is a complex parameter. 

Other than these soft masses, it is necessary to know the tree level 
Higgs soft masses
\begin{eqnarray}
M_{H_u}^{2}=y_u |\mu^{2}|~,\ \ \ M_{H_d}^{2}=y_d |\mu ^{2}|~,\ \ \
\mu\ B=|\mu|^{2} (\frac{8 m_s^{2}}{v_{s}^{2}}+k_{\mu})~,
\end{eqnarray}
where $m_s^2\sim v_s^2$ is a natural choice as discussed in Ref.~ \cite{12}.  Here $y_u$ and $y_d$ are 
real parameters, and $k_{\mu}$ is a complex parameter determining the phase of 
the $B$ parameter. As was analyzed in Ref.~11 in  detail this phase 
can be identified with the relative phase of the Higgs doublets; hence, there 
is no CP violation in Higgs sector of (\ref{softMSSM}) at tree level.

After electroweak breaking the Higgs doublets in (\ref{softMSSM}) can be expanded as 
\begin{eqnarray}
\label{doublet}
H_{d}&=&\left(\begin{array}{c c} H_{d}^{0}\\
H_{d}^{-}\end{array}\right)=\frac{1}{\sqrt{2}}\left(\begin{array}{c c}
v_{d}+\phi_{1}+i\varphi_{1}\\ H_{d}^{-}\end{array}\right)\;,\nonumber\\
H_{u}&=&\left(\begin{array}{c c} H_{u}^{+}\\
H_{u}^{0}\end{array}\right)=\frac{e^{i\theta}}{\sqrt{2}}\left(\begin{array}{c c}
H_{u}^{+}\\ v_{u}+\phi_{2}+i\varphi_{2}\end{array}\right)\; . 
\end{eqnarray}
where $\tan\beta\equiv v_u/v_d$ as usual, and the angle parameter $\theta$ is 
the misalignment between the two Higgs doublets. As in Ref.~11 the 
angle $\theta$ gets embedded into the total CP violation angle $\mbox{Arg}[\mu A_t]$,
and we will not elaborate radiative corrections to it \cite{28}.

As usual, we calculate the Higgs masses and their mixings up to one loop accuracy via
\begin{eqnarray}
M^{2}=\left(\frac{\partial^{2}\ V} {\partial \chi_{i} \partial \chi_{j}}\right)_{0}\,,
\mbox{where}\; 
\chi_{i} \in {\cal{B}}=\{\phi_{1}, \phi_{2}, \varphi_{1}, \varphi_{2}\} \; .
\end{eqnarray}
where $V\equiv V_0 + V_{1-loop}$ is the radiatively corrected Higgs
potential \cite{11}.
As mentioned before, we take into account only top quark and top squark loop corrections, which are 
the dominant ones as long as $\tan\beta\simlt 50$. The radiative corrections depend on the stop 
mass-squared eigenvalues $m_{\tilde{t}_{1,2}}^{2}$ 
\begin{eqnarray}
\label{light}
m_{\tilde{t}_{1,2}}^{2}=\frac{1}{2}\left ((k_{u}^{2}+ k_{Q}^{2})|\mu|^{2}
+ 2 m_{t}^{2} \mp \Delta_{\tilde{t}}^{2}\right),
\end{eqnarray}
whose splitting 
\begin{eqnarray}
\label{del}
\Delta_{\tilde{t}}^{2}=|\mu|\sqrt{ (k_{u}^{2}- k_{Q}^{2})^{2}\ |\mu|^{2}+
4 m_{t}^{2}(|k_{t}|^{2}+\cot^{2} \beta-2 |k_{t}| \cot \beta \cos \varphi_{kt})}\; .
\end{eqnarray}
will play a key r{\^o}le in analyzing the results as it depends explicitly on the total 
CP violation angle 
\begin{eqnarray}
\varphi_{kt}=\mbox{Arg}[\mu A_{t}]=\mbox{Arg}[k_t]
\end{eqnarray}
where $k_{t}$ has been defined in (\ref{at}). One here notices that $\Delta_{\tilde{t}}^{2}$ increases
as $\varphi_{kt}$ changes from 0 to $\pi$. This particularly means that the strength of the radiative
corrections are modified as $\varphi_{kt}$ ranges from one CP--conserving point to the next.

We express the (3$\times$3) dimensional Higgs mass--squared matrix 
\begin{eqnarray}
\label{massmat}
M^{2}=\left(\begin{array}{c c c}
M_{11}+\Delta M_{11} & M_{12}+\Delta M_{12} & \Delta M_{13}\\
M_{12}+\Delta M_{12} & M_{22}+\Delta M_{22} &\Delta M_{23}\\
\Delta M_{13} & \Delta M_{23} &M_{33}+\Delta M_{33}\\
\end{array}\right)~,
\end{eqnarray}
in the basis ${\cal{B}}=\{\phi_{1}, \phi_{2},
\sin\beta \varphi_{1}+\cos\beta \varphi_{2}\}$  using (\ref{doublet}). 
The elements  of the mass matrix read as below:
\begin{eqnarray}
\label{m11}
M_{11}&=& M_{Z}^{2} \cos^{2}\beta + \tilde{M}_{A}^{2}
\sin^{2}\beta~,\nonumber\\
M_{12}&=&-(M_{Z}^{2}+\tilde{M}_{A}^{2})\sin\beta\cos\beta~,\nonumber\\
M_{22}&=&M_{Z}^{2} \sin^{2} \beta  +\tilde{M}_{A}^{2}cos^{2}\beta~, \nonumber\\
M_{33}&=&\tilde{M}_{A}^{2}~,
\end{eqnarray}
where the radiative corrections are generically denoted by $\Delta M_{ij}$.
In the case of CP-conserving limit in which 
CP-even and CP-odd sectors in (\ref{massmat}) decouple,
$\tilde{M}_{A}^{2}$ becomes the radiatively corrected pseudoscalar mass
and $\Delta M_{11, 12, 22}$ become the usual one-loop corrections to the CP
even scalar mass -squared matrix \cite{18,19}.
On the other hand,  $\Delta M_{13}$
and  $\Delta M_{23}$  are genuiely generated by the SUSY
CP--violation effects. 
We define
\begin{eqnarray}
{\cal{ R}}_{kt}&=&|k_{t}|\cos\varphi_{kt} - \cot\beta~,\nonumber\\
{\cal{ L}}_{kt}&=&|k_{t}| - \cot\beta\cos\varphi_{kt}~,\nonumber\\
{\cal{ C}}_{kt}&=&(|k_{t}|-\cot\beta)\sin^{2} \varphi_{kt}~.
\end{eqnarray}
These  radiative  correction terms have the following expressions:
\begin{eqnarray}
\Delta M_{11}& = &-2\beta_{h_{t}}|\mu|^{4}m_{t}^{2}
\frac{{\cal{ R}}_{kt}^{2}}{\Delta_{\tilde{t}}^{4}}
g(m_{\tilde{t}_{1}}^{2},m_{\tilde{t}_{2}}^{2})~,
\end{eqnarray}
\begin{eqnarray}
\Delta M_{12}& =& -2\beta_{h_{t}}m_{t}^{2}|\mu|^{2}
\big[\frac{ {\cal{R}}_{kt}}{ {\Delta_{\tilde{t}}^{2}}} 
\log{\frac{m_{\tilde{t}_{2}}^{2}}{m_{\tilde{t}_{1}}^{2}}} \nonumber\\
&-&|\mu|^{2}|k_{t}|\frac{ {\cal{R}}_{kt}^{2}+|k_{t}|{\cal{ C}}_{kt}} 
{\Delta_{\tilde{t}}^{4}}g(m_{\tilde{t}_{1}}^{2},m_{\tilde{t}_{2}})\big]~,
\end{eqnarray}
\begin{eqnarray}
\Delta M_{22}& =& 2 \beta_{h_{t}}  m_{t}^{2}\big[\log {
\frac{m_{\tilde{t}_{2}}^{2}
m_{\tilde{t}_{1}}^{2}}{m_{t}^{4}}}+
 2 |k_{t}||\mu|^{2} \frac{{\cal{ L}}_{kt}}{\Delta_{\tilde{t}}^{2}}
\log{\frac{m_{\tilde{t}_{2}}^{2}}{m_{\tilde{t}_{1}}^{2}}}\nonumber\\
&-&|k_{t}|^{2}|\mu|^{4}
\frac{{\cal{ L}}_{kt}^{2}}{\Delta_{\tilde{t}}^{4}}
g(m_{\tilde{t}_{1}}^{2},
m_{\tilde{t}_{2}}^{2})\big]~,
\end{eqnarray}
\begin{eqnarray}
\Delta M_{13}&=& -2 \beta_{h_{t}}m_{t}^{2}|\mu|^{4}|k_{t}|
\frac{\sin\varphi_{kt}}{\sin\beta}
\frac{{\cal{ R}}_{kt}}{\Delta_{\tilde{t}}^{4}}
g(m_{\tilde{t}_{1}}^{2},m_{\tilde{t}_{2}}^{2})~,
\end{eqnarray}
\begin{eqnarray}
\Delta M_{23}& =& -2 \beta_{h_{t}}m_{t}^{2}\frac{|\mu|^{4}
|k_{t}|^{2}}{\Delta_{\tilde{t}}^{4}}
\frac{\sin\varphi_{kt}}{\sin\beta}\nonumber\\&\times&
\big[{\cal{ L}}_{kt}-\frac{1}{|k_{t}||\mu|^{2}
g(m_{\tilde{t}_{1}}^{2},m_{\tilde{t}_{2}}^{2})}\Delta_{\tilde{t}}^{2}
\log{\frac{m_{\tilde{t}_{2}}^{2}}{m_{\tilde{t}_{1}}^{2}}}\big]~,
\end{eqnarray}
\begin{eqnarray}
\Delta M_{33}& =& -2 \beta_{h_{t}}m_{t}^{2}
\frac{\sin\varphi_{kt}^{2}}{\sin\beta^{2}}
\frac{|\mu|^{4}|k_{t}|^{2}}{\Delta_{\tilde{t}}^{4}}
g(m_{\tilde{t}_{1}}^{2},m_{\tilde{t}_{2}}^{2})~,
\end{eqnarray}
where $\beta_{h_{t}}=3h_{t}^{2}/16\pi ^{2}$,
and  the function $g(x,y)$ is defined by
\begin{eqnarray}
g(m_{\tilde{t}_{1}}^{2},m_{\tilde{t}_{2}}^{2})&=&
-2+\frac{ m_{\tilde{t}_{2}}^{2} + m_{\tilde{t}_{1}}^{2}}
{m_{\tilde{t}_{2}}^{2}  - m_{\tilde{t}_{1}}^{2}}
\log{\frac{m_{\tilde{t}_{2}}^{2}}{m_{\tilde{t}_{1}}^{2}}}.
\end{eqnarray}

We diagonalize the Higgs mass--squared matrix (\ref{massmat}) by the similarity 
transformation
\begin{eqnarray}
{\cal{R}}M^{2}{\cal{R}}^{T}= {\rm diag}(m_{h_{1}}^{2},
m_{h_{2}}^{2}, m_{h_{3}}^{2})~,
\end{eqnarray}
where  ${\cal{R}}{\cal{R}}^{T}=1$.  In
the following, we define $h_3$ to be the lightest of all three;
$h_2$ to be the one that corresponds to the heavy pseudoscalar Higgs boson and $h_1$
to be  the heavy CP-even scalar Higgs boson.
One of the most important quantities in our analyses is 
the percentage CP composition of a given mass--eigenstate Higgs boson.
The percentage  CP compositions of the Higgs bosons in terms of the basis
elements are defined by 
\begin{eqnarray}
\rho_{i}=100\times |{\cal{R}}_{1i}|^{2};\,\,\, i=1, 2, 3.
\end{eqnarray}
where $\rho_{1}$, $\rho_{2}$  and $\rho_{3}$ correspond respectively
the $\phi_{1}$, $\phi_{2}$,
$\sin\beta \varphi_{1}+\cos\beta \varphi_{2}$  components of the Higgs boson
under concern. 

In what follows, we will make the numerical analysis for 
evaluating the masses of the Higgs scalars and 
analyzing the relative strengts of their percentage CP compositions under the effects of SUSY CP
phases. In doing this, we will first focus on the 
percentage CP compositions of the  
lightest Higgs boson ($h_3$), especially  its CP--odd composition 
($\rho_{3}$), which can offer new opportunities at colliders for observing the
Higgs boson \cite{8,29}. We will discuss the dependence of the  CP odd compositions of
$h_3$  on the  CP--breaking angle  
in reference to the previous theoretical \cite{8,11} as well as the recent 
experimental bounds \cite{16}. Next, we will analyze the masses and the
percentage CP compositions of the 
remaining  two heavy scalars  ($h_1$, $h_2$) in the low and high $\tan\beta$ regimes.

\section{Numerical Analysis}
Using the formulae in the last section, we will now analyze several quantities
in a wide range of the parameter space. As a reflecting property of the
model, all parameters are expressed in terms of the  $\mu$ parameter. Since the $\mu$ parameter is already
stabilized to the weak scale, as a consequence of the naturalness, all dimensionless
quantities are expected to be ${\cal{O}}(1)$. Therefore, as a representative point
in the parameter space we take 

\begin{eqnarray}
k_Q=k_u=|k_t|=1~,
\end{eqnarray}

In our analysis,  $|\mu|$ changes from $250~{GeV}$ to $1000~{GeV}$ and 
$M_{A}$  from  $|\mu|$ to $5\ |\mu|$ for each $\mu$ value
in the full $\varphi_{kt}$ range.
However, we would like to note that for
the values of $\mu \simlt 450~{GeV}$, it is not possible to find regions in
the parameter space satisfying the recent LEP  constraints \cite{16}.  
Moreover, we  concentrate on the two specific values of 
$\tan\beta$ namely $\tan\beta=4$ and $\tan\beta=30$
to analyze the behaviour of the Higgs masses and mixings in the  small and large 
$\tan\beta$ regimes in detail. 

\begin{figure}[htb]
\centerline{\epsfig{file=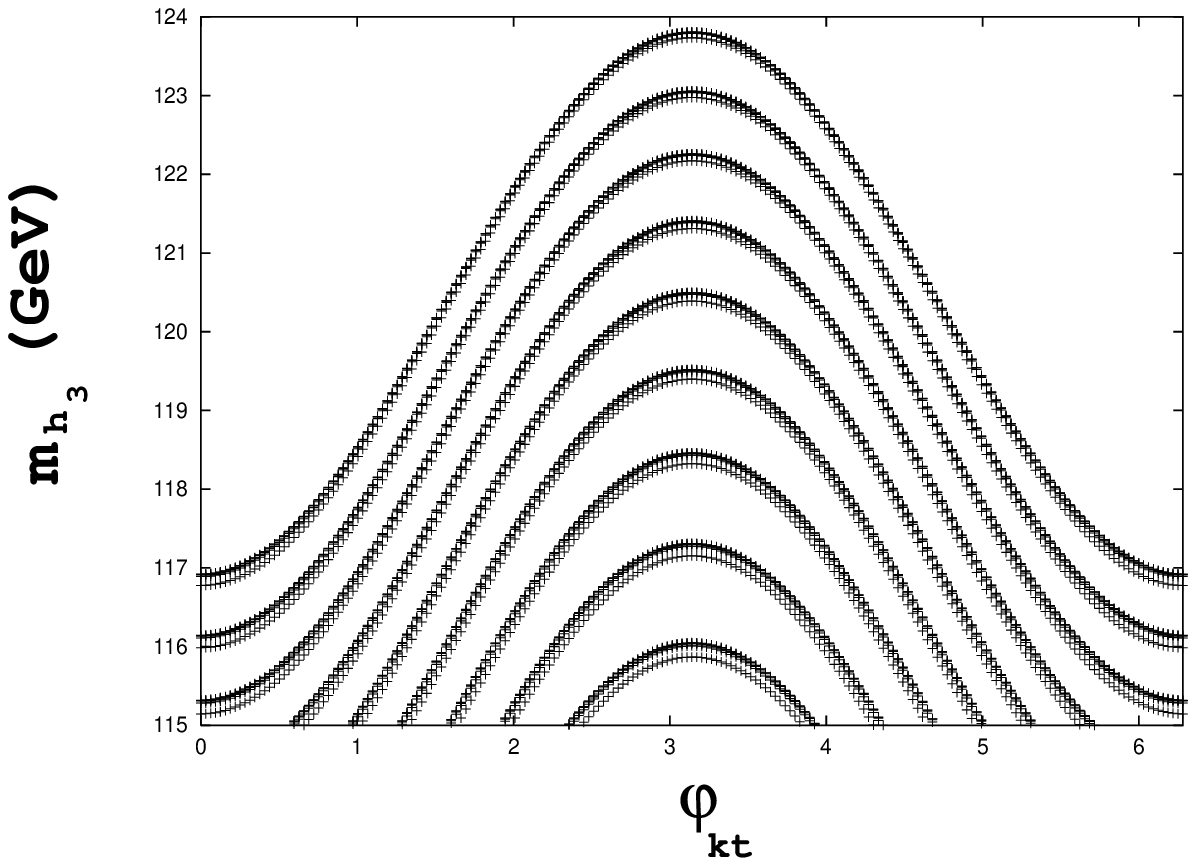,height=2.6in,width=2.6in }
\epsfig{file=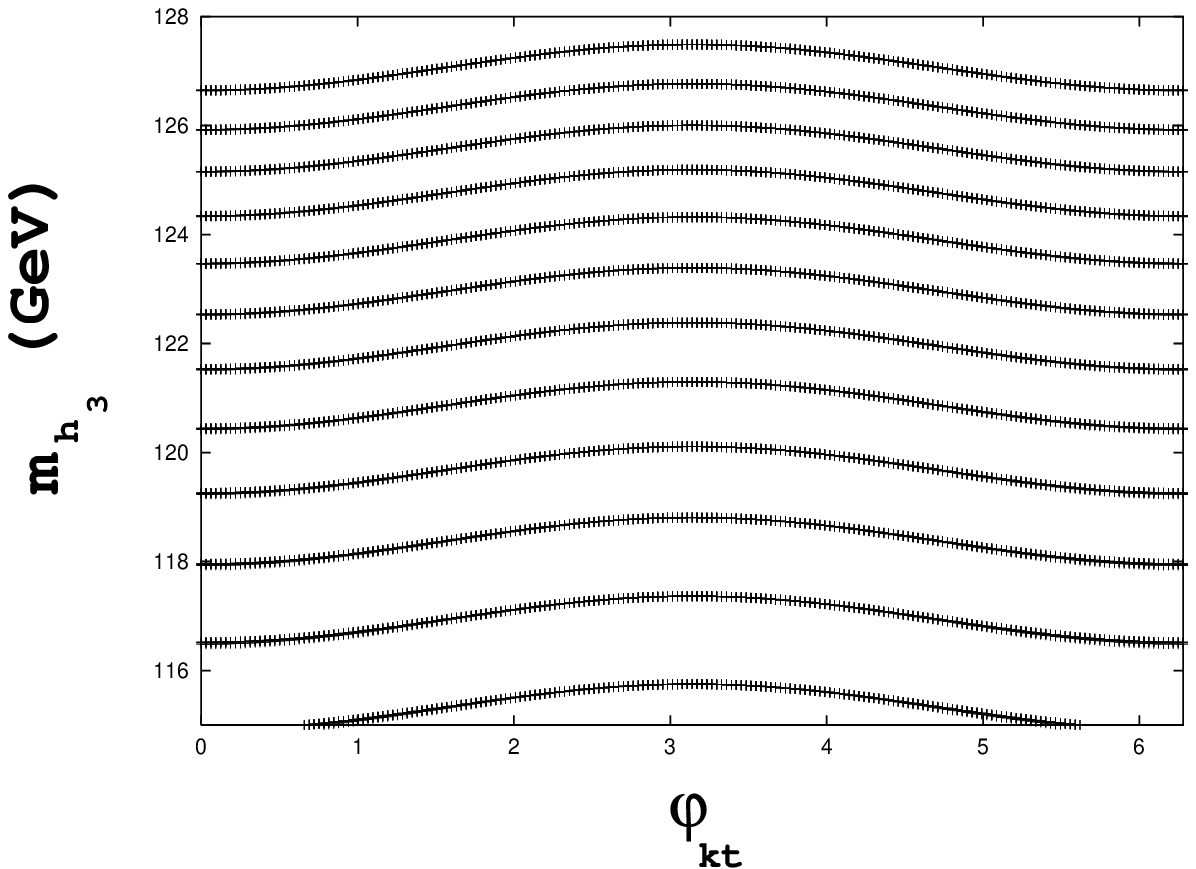,height=2.6in,width=2.6in }}
\caption{ The mass ($m_{h_{3} }$) of the lightest Higgs boson ($h_{3}$),
as a function of $\varphi_{kt}$ 
for  $\tan\beta$=4 (left panel) and $\tan\beta$=30 (right panel).}
\end{figure}

Fig. 1  illustrates the dependence of the lightest Higgs mass $m_{h_{3}}$  on
$\varphi_{kt}$ for $\tan\beta=4$ (left panel) and $\tan\beta=30$ (right
panel). 
One immediately observes that $m_{h_{3}}$ increases with $\varphi_{kt}$
for $\tan\beta=4$ (left panel), whereas it remains nearly constant for $\tan\beta=30$
(right panel) in the $[0,\pi]$ interval.
This saturation effect in the Higgs mass can be easily understood
by observing that the radiative corrections depend strongly on the
stop splitting $\Delta_{\tilde{t}}^{2}$. This quantity depends explicitly
on $\varphi_{kt}$ such that: 

\begin{eqnarray}
\label{ratio}
\frac{{\Delta_{\tilde{t}}^{2}(\pi)}}
{{\Delta_{\tilde{t}}^{2}(0)}}\sim \sqrt{\frac{1+\sin2 \beta}{1-\sin2 \beta}}~,
\end{eqnarray}
that is, the stop splitting   $\Delta_{\tilde{t}}^{2}$ increases with
increasing $\varphi_{kt}$,
as  $\varphi_{kt}$ changes from 0 to $\pi$.
This particularly  implies that the strength of the radiative corrections modify as   $\varphi_{kt}$
changes from one CP-conserving point to the next.
However, (\ref{ratio}) decreases with increasing $\tan\beta$. Indeed, it approaches to unity in the large 
$\tan\beta$ limit.  Therefore, the radiative corrections to the lightest
Higgs mass $m_{h_{3}}$ which 
are sensitive to variations in $\varphi_{kt}$ are suppressed in large 
$\tan\beta$ regime. In the light of these observations, it is clear that
the lightest Higgs mass ${m_{h}}_{3}$ is much more flat for $\tan\beta=30$ (right panel) compared to that 
for $\tan\beta=4$ (left panel).   

As the left panel of Fig. 1 suggests that,
the  $\varphi_{kt}$ dependence
of $m_{h_{3}}$ around  $(\phi_{kt}=\pi)$
differs from those at other CP-conserving points, in particular for the case
of $\tan\beta=4$.
That is, 
the maximal value of the lightest Higgs mass $m_{h_{3}}$ occurs at $(\phi_{kt}=\pi)$, and 
then, the radiative corrections
reverse their sign as $\varphi_{kt}$ changes  from  $\pi$ to  $ 2 \pi$ 
(see (\ref{light}), (\ref{del})).
Numerically,  for $\tan\beta=4$,  $m_{h_{3}}(\phi_{kt}=\pi)$
is larger than $m_{h}(\phi_{kt}=0)$ by $\sim 10~{\rm GeV}$ due to
the enhancements in the radiative corrections as $\varphi_{kt}$
ranges from 0 to $\pi$. 
In contrast to $\tan\beta=4$ case, the sensitivity
of the lightest Higgs mass $m_{h_{3}}$  on  $\varphi_{kt}$ is washed out in the large $\tan\beta$
regime due to the reasons explained above (\ref{ratio}).

It is known that the recent experimental data requires   $m_{h_{3}}\simgt 115~{GeV}$\cite{16}.
Imposing this constraint on $m_{h_{3}}$, the experimental bound on the lightest 
Higgs mass is satisfied  for all the parameter space, for which
$\tan\beta=4$ and  $\tan\beta=30$.
The dependence of the Higgs mass on the CP violation
angle has also been noted in Ref.~8, for general minimal supersymmetric
standart model with a limited range of the parameters.

\begin{figure}[htb]
\centerline{\epsfig{file=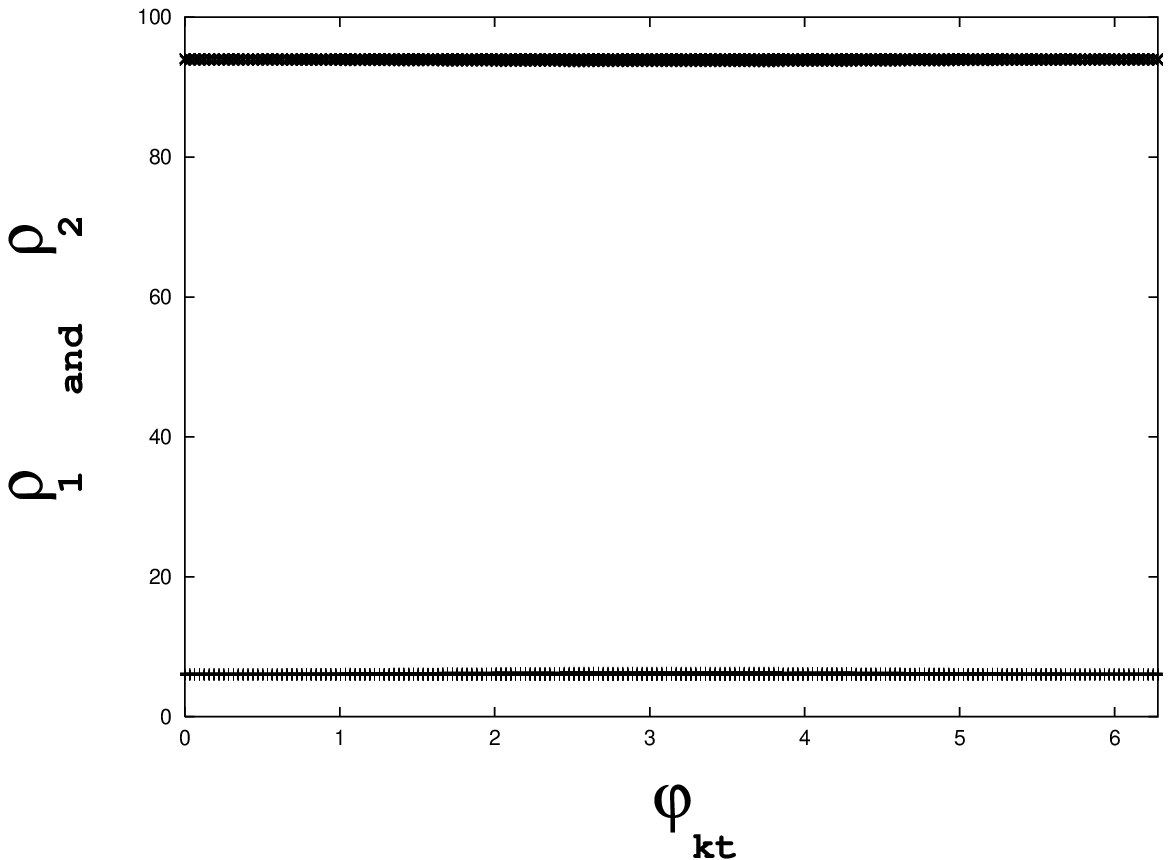,height=2.6in,width=2.6in }
\epsfig{file=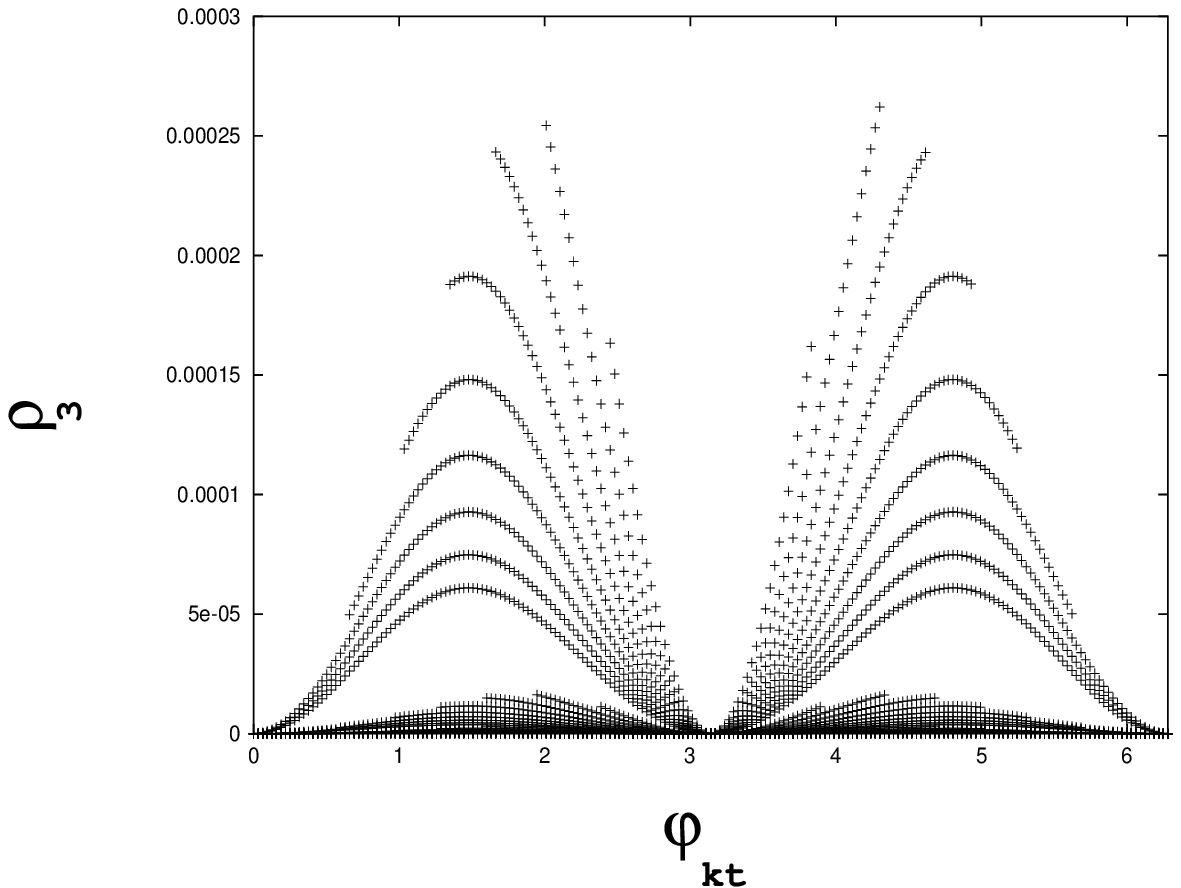,height=2.6in,width=2.6in }}
\caption{ The percentage CP--even
compositions  of the lightest Higgs
boson $h_{3}$ (left panel) where  the bottom  and the top  curves present $\rho_{1}$
and  $\rho_{2}$, respectively, and its percentage CP--odd composition
$\rho_{3}$ (right panel),  as a function of $\varphi_{kt}$ for  $\tan\beta$=30.}
\end{figure}

In Fig. 2 ,  we show the $\varphi_{kt}$ dependence of
the percentage CP-even, $\rho_{1}$--$\rho_{2}$, (left panel),  and the
percentage CP--odd  $\rho_{3}$ (right panel) compositions of the lightest
Higgs boson ($h_{3}$),  for $\tan\beta=4$.
As is seen from the left  panel  Fig. 2,
$h_{3}$  has $\approx 94\%$ $\rho_{2}$ and $\approx 6\%$  $\rho_{1}$
compositions for $\tan\beta=4$. On the other hand,  the  right panel of Fig. 2 
suggests that its percentage CP--odd
composition  $\rho_{3}$ is extremely small for  small  
$\tan\beta$ regime.  Numerically, the maximum value of $\rho_{3}$ is  $\approx 0.0003\%$  
in the full range of  $\varphi_{kt}$ for $\tan\beta=4$ (right panel).

Depicted in  Fig. 3 is the percentage composition  CP-even, 
$\rho_{1}$--$\rho_{2}$, (left panel),
and percentage CP--odd  $\rho_{3}$ (right panel) compositions of the lightest
Higgs boson ($h_{3}$),  for $\tan\beta=30$.
One notes that   $\rho_{2}$  increases near to the
$\approx 99.9\%$,  while $\rho_{1}$ remains below $\approx 0.12\%$ for $\tan\beta=30$.
On the other hand, as is seen from the right panel of Fig. 3,  
the  percentage CP--odd  composition  ($\rho_{3}$) of the lightest
Higgs boson ($h_{3}$)  is still very small. However,
it  increases relatively as compared to that for
$\tan\beta=4$ (left panel), and  reaches to a maximum
value of $\approx 0.0013\%$ for $\tan\beta=30$  in the entire range of  $\varphi_{kt}$. 

One notes that  the CP--odd component  of $h_{3}$ 
never exceeds  $0.0013\%$ in the full  $\varphi_{kt}$ range, for all values
of $\tan\beta$ changing from 4 to 30.  
Compared to its CP--even compositions, which form the remaining percentage, 
this CP--odd component is extremely small to cause observable effects. 
It may, however, be still important when the radiative 
corrections to gauge and Higgs boson vertices are included \cite{8}.

\begin{figure}[htb]
\centerline{\epsfig{file=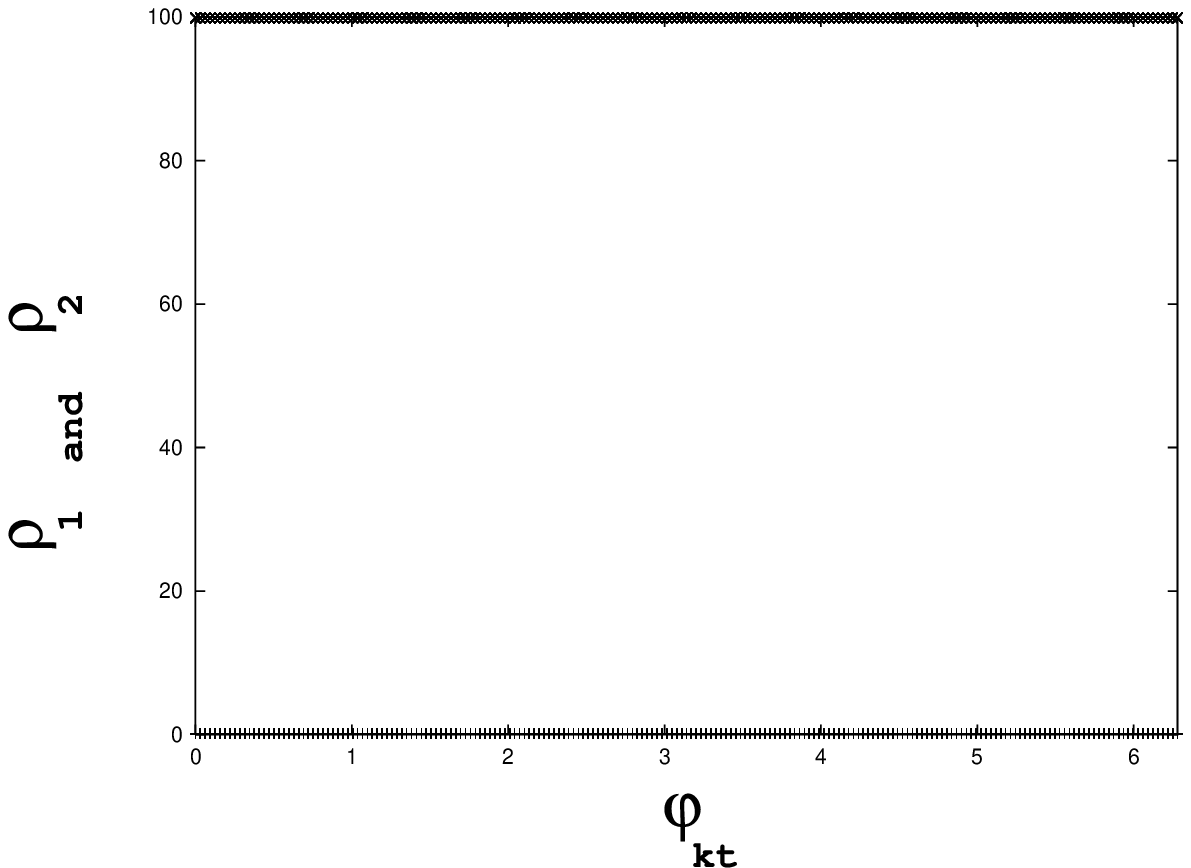,height=2.6in,width=2.6in }
\epsfig{file=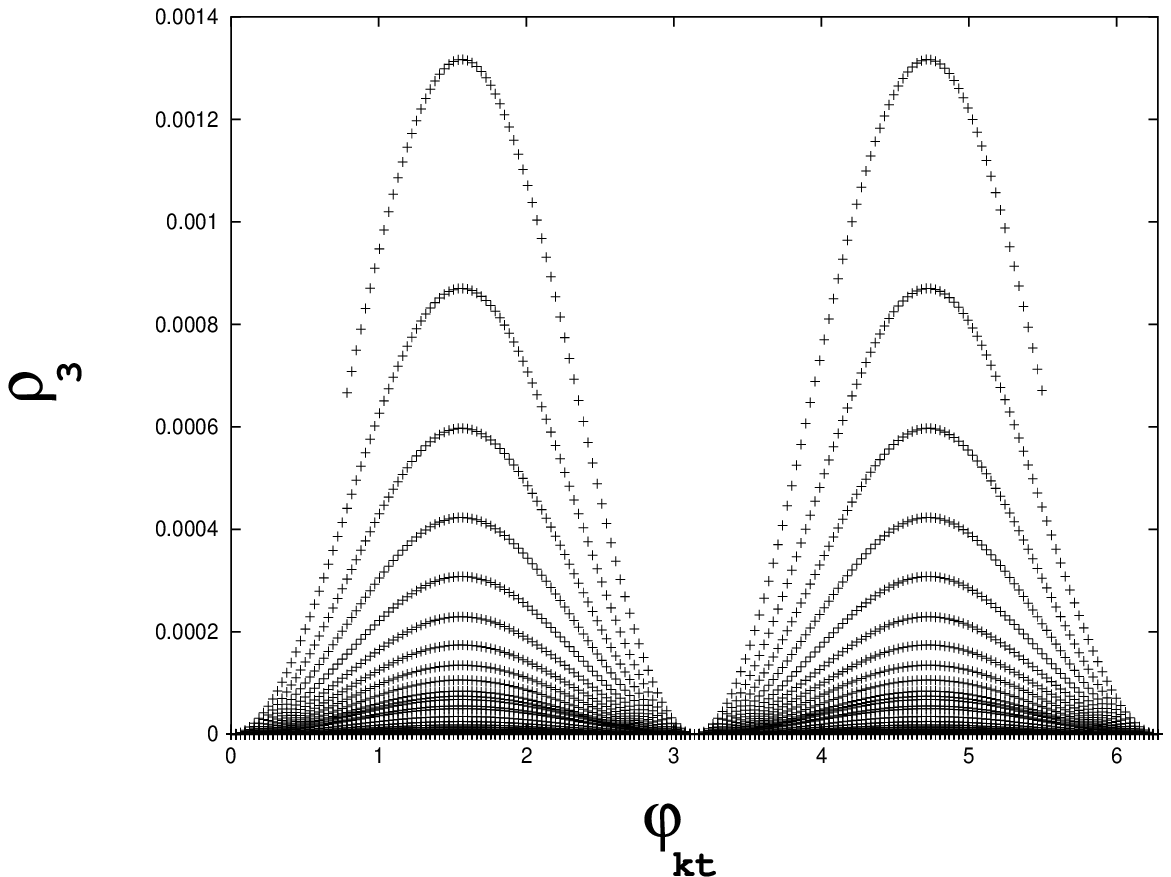,height=2.6in,width=2.6in }}
\caption{ The percentage CP--even
compositions  of the lightest Higgs
boson $h_{3}$ (left panel) where  the bottom  and the top  curves present $\rho_{1}$
and  $\rho_{2}$, respectively, and its percentage CP--odd composition
$\rho_{3}$ (right panel),  as a function of $\varphi_{kt}$ for  $\tan\beta$=30.}
\end{figure}

Depicted  Fig. 4 is the $|\mu|$ dependence
of $\rho_3$  designating the CP--odd percentage composition of $h_{3}$, for
$\tan\beta=4$(left panel), and  $\tan\beta=30$(right  panel), respectively.  
Both the left and right panels of Fig. 4 suggest that  the  CP--odd percentage composition 
of $h_{3}$ ($\rho_3$) decreases with $|\mu|$. 
It is also seen from the  left panel of Fig. 4 that , the maximum value of $\rho_{3}$
($\approx 0.00027\%$)
occurs at  $|\mu|\approx 650~{GeV}$.
For larger values of $|\mu|$,  $\rho_3$ decreases gradually as the
supersymmetric spectrum decouples. For smaller values of $|\mu|$, however, 
the parameter space is constrained by the existing LEP bound on the lightest
Higgs mass \cite{16}. That is,  the CP--odd percentage composition of $h_{3}$ ($\rho_3$) gets smaller until $\mu \approx 600~{GeV}$ and 
it is not possible to find any region in the parameter space
below this value ($\mu \simlt 600~{GeV}$) for $\tan\beta=4$ (left panel), since this region is completely disallowed 
by the experimental bound \cite{16}.
On the other hand, as is shown in the  right panel of Fig. 4,  that  the maximum value of 
the CP--odd percentage composition of $h_{3}$ 
($\rho_3 \approx 0.0013\%$) occurs at  $|\mu|\approx 450~{GeV}$ for $\tan\beta=30$.
As is in the left panel of Fig. 4, 
it again decreases with increasing $|\mu|$. 
Therefore (remembering  $M_{A}\propto |\mu|$ for the model under concern) unless
$|\mu|$ is choosen smaller (equivalently unless the Peccei--Quinn scale 
is pushed down towards the lower limit of the allowed $axion$ $window$
\cite{12,15}) one cannot increase the CP--odd composition of the lightest Higgs boson.

\begin{figure}[ht]
\centerline{\epsfig{file=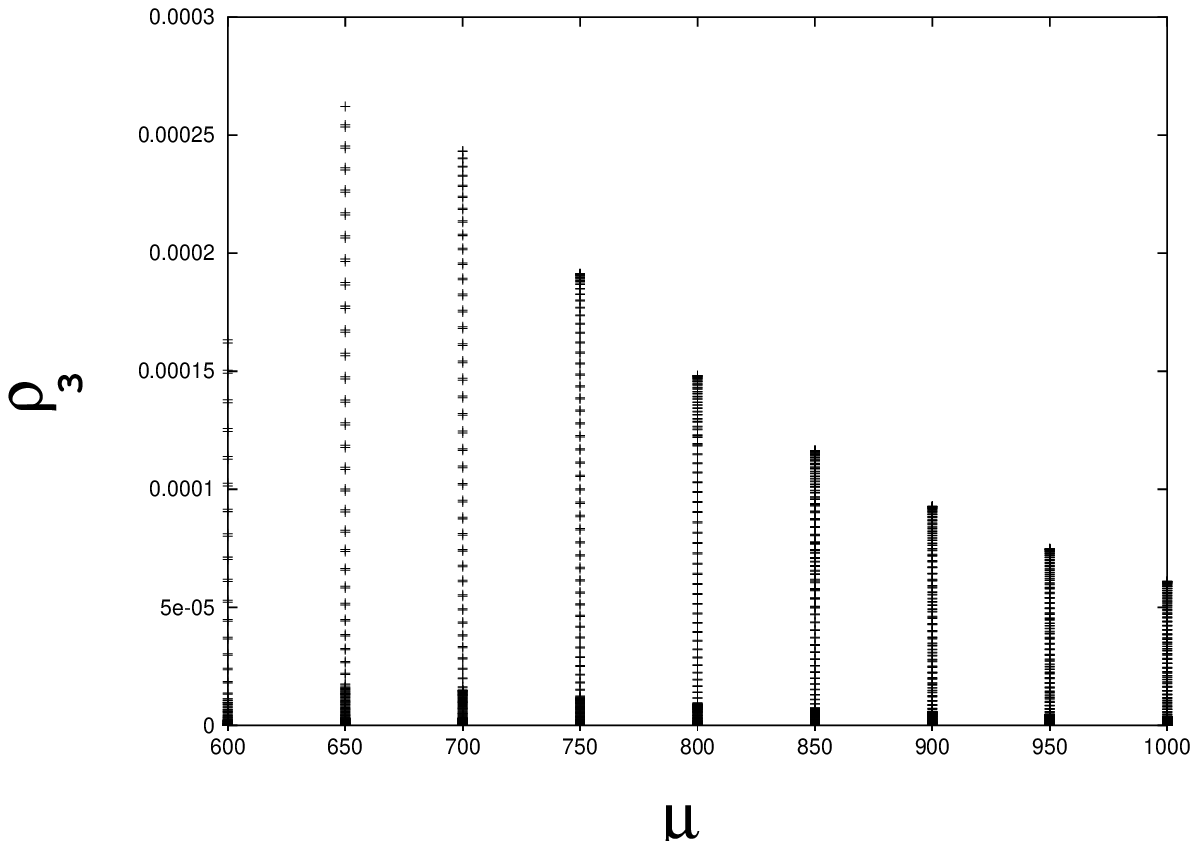,height=2.6in,width=2.6in }
\epsfig{file=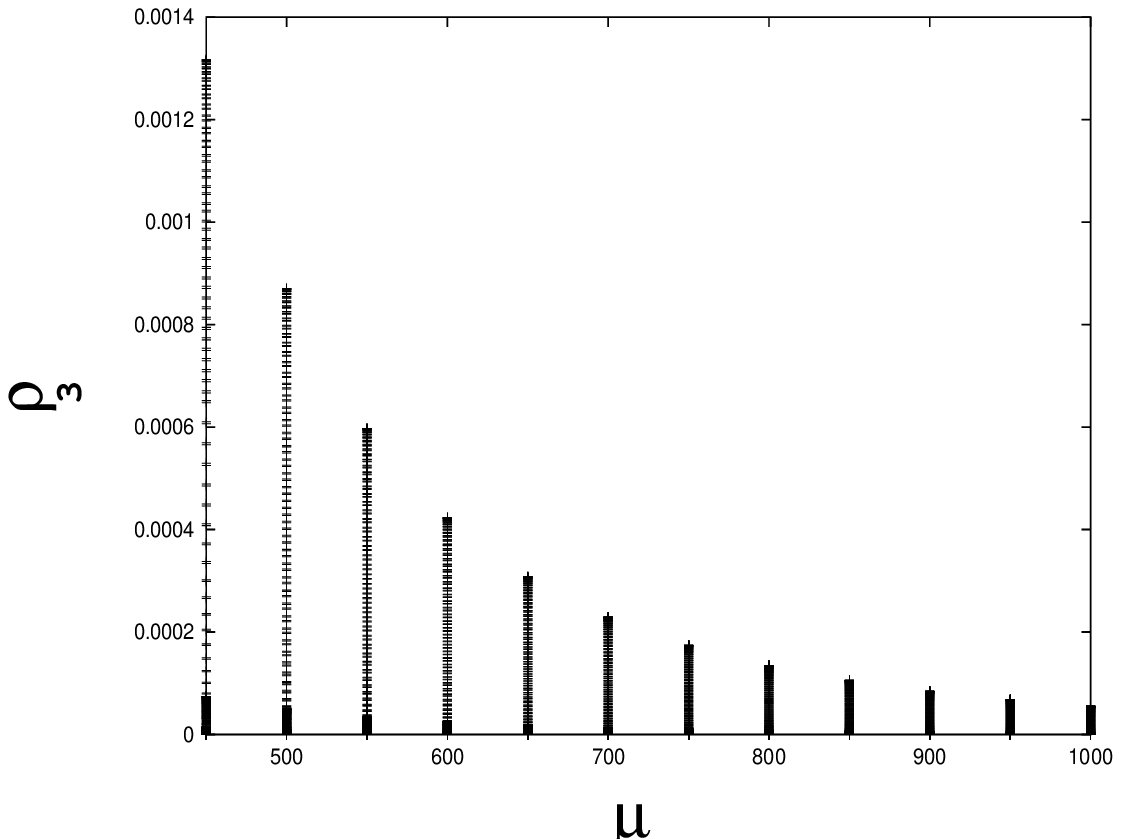,height=2.6in,width=2.6in }}
\caption{ The CP--odd composition ($\rho_{3}$) of the lightest Higgs ($h_{3}$),
as a function of $\mu$  for  $\tan\beta=4$(left panel), 
 and $\tan\beta$=30 (right panel).}
\end{figure}

One notes that, the lower limit on the lightest  Higgs mass is directly
correlated with its CP--odd composition, that is, as the lower bound on the
lightest Higgs mass increases, its CP--odd composition decreases as will be
indicated by Fig. 5  

In Fig. 5, we show the variation of the lightest Higgs mass ($m_{h_{3}}$) with its
CP--odd composition ($\rho_3$) for $\tan\beta=4$(left panel), $\tan\beta=30$
(right  panel), respectively.
Both windows  of the figure suggest that, lighter the Higgs boson, ($m_{h_{3}}$), larger its CP odd
composition ($\rho_3$).
From the left panel of Fig. 5, it is seen that the maximum value of the  CP--odd
composition of $h_{3}$ starts from  $\approx 0.00027\%$
and decreases rapidly for  $\tan\beta=4$ (left panel).
On the other hand, as the right panel of the figure suggests that the maximum value of  its CP--odd composition occurs at  
$\rho_3 \approx 0.0013\%$  for
$\tan\beta=30$, and again relatively decreasing  with the increasing mass,
it reaches far below  $0.0002\%$ for $m_{h_{3}}= 127~{GeV}$.

One also notes that,  the same kind of variation can be observed for all values of
$\tan\beta$, ranging from 4 to 30. 
Therefore, one can conclude that  possible increase in the lower experimental bound of the
lightest Higgs mass in  future  colliders  will imply reduced  CP--odd composition.

\begin{figure}[htb]
\centerline{\epsfig{file=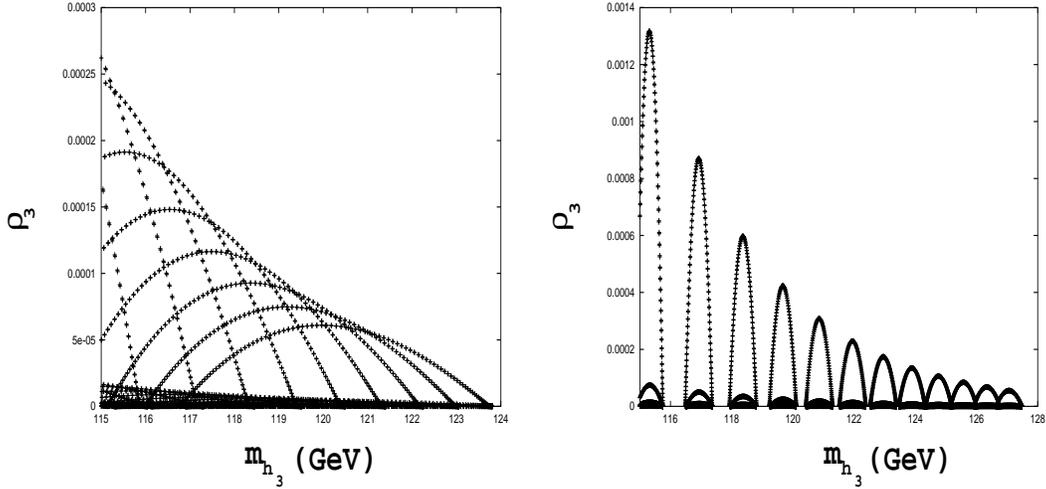,height=2.6in,width=5.5in}}
\vspace*{0.1truein}
\caption{ The variation of 
the mass ($m_{h_{3} }$) of the lightest Higgs boson ($h_{3}$),
with its  CP--odd composition ($\rho_{3}$)   
for  $\tan\beta$=4 (left panel) and $\tan\beta$=30 (right panel).}
\end{figure}

In the first part of our numerical analysis, we have studied  
the percentage CP compositions of the lightest Higgs boson $h_{3}$, in
particular its CP--odd composition,  as well as its mass 
for a given portion of the parameter space which are of prime importance
in the light of present LEP experiments\cite{16}. On the other hand,   
it is a well--known fact that the heavy Higgs bosons, which are out of
reach of the present colliders, have no definite CP quantum numbers for most
of the MSSM parameter space and  it will be hard
to observe them before NLC or TESLA operates. However, at this point, we also would like to
discuss CP  characteristics of the heavy Higgs bosons for the underlying
model, in the low and high $\tan\beta$  
regimes, for completeness.

\begin{figure}[htb]
\centerline{\epsfig{file=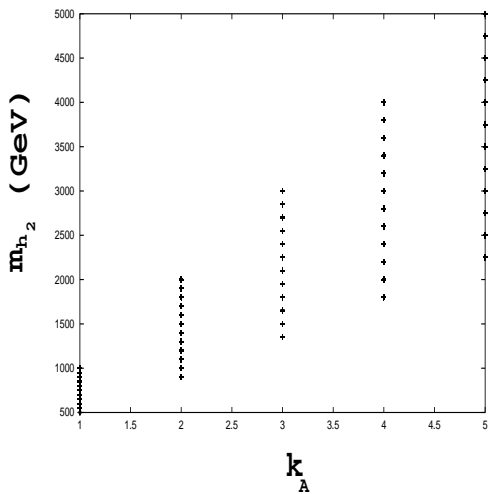,height=2.6in,width=2.6in }}
\caption{The mass   of the second heavy scalar ($m_{h_{2}}$) as a function of   
 $k_{A}$ for all values of $\tan\beta$ changing from  4 to 30.}
\end{figure}

In Fig. 6, we show the variation of  the mass of the second heavy scalar
$m_{h_{2}}$ with  $k_A$, for all the values of $\tan\beta$ ranging from 4 to 30.
Since  the two heavy scalars are degenerate in mass,    
$k_A$ dependence of  $m_{h_{2}}$ is indicated in Fig. 5, for convenience.
It is seen from the figure that  the lower bound of $m_{h_{2}}$  starts from $500~{\rm GeV}$, and
extends to $5000~{\rm GeV}$, while $M_{A}$ ranges from  $|\mu|$ to $5\ |\mu|$ for each $\mu$ value.
One notes that, for $k_A=1$ the masses of the heavy scalars changes from
$500~{\rm GeV}$ to $1000~{\rm GeV}$, lying right at the weak scale.

Depicted in  Fig. 7, is  the  $\varphi_{kt}$ dependence of 
$\rho_{1}$--$\rho_{3}$ designating  
the  CP--even and CP--odd percentage compositions of the second scalar particle
($h_{2}$) respectively, for  $\tan\beta=4$ (left panel) and
$\tan\beta=30$ (right panel).
As  is  noticed from the figure  that as the $\rho_{3}$ component of $h_{2}$
changes  in between $\approx 100\% $ and   $\approx 98.2\%$, 
its  $\rho_{1}$ component becomes at most  $\approx 1.8\%$ for
$\tan\beta=4$ (left panel).
On the other hand,  in passing to the large 
$\tan\beta$ regime, one notes that 
there is a complementary  behaviour of $\rho_{3}$ (right panel).
Starting from $\varphi_{kt}=0$ at the $100\%$ level, it vanishes at
$\varphi_{kt}=\pi/2$ at the $ 0\% $ level, then it increases to $ 100\% $ at
$\varphi_{kt}=\pi$,  decreasing to  $0\%$ level again at
$\varphi_{kt}=3\pi/2$,
it  completes its behaviour at the  $100\% $ level.  
Its  $\rho_{1}$ component follows the same behaviour but it starts from
$\varphi_{kt}=0$ at the $ 0\%$ level. 
It is seen that  though  the CP--conserving
points, namely $\varphi_{kt}=0,\pi, 2\pi$,
the particle under concern has a definite CP-parity. On the other hand, at the maximal
CP--violation points,  namely $\varphi_{kt}=\pi/2,3\pi/2$, the CP--parity of
the particle is completely reversed. And
except for the points mentioned above, the particle has no definite CP characteristics. 
In summary, $h_{2}$ is a pure pseudoscalar for small $\tan\beta$ values,
whereas its CP-parity swings significantly as the CP-phase varies
in the large $\tan\beta$ regime   

As  seen from Fig. 8,  $h_{1}$
has $\approx 94\% $  $\rho_{1}$  and $\approx 1.8\%$ $\rho_{3}$
composition for $\tan\beta=4$ (left panel) and the particle under concern is
a  CP-even scalar.
In the large   $\tan\beta$ regime however, in accordance with the right
panel of Fig. 7,  $h_{1}$  has no
definite CP parity except for the pure CP--conserving and CP--violating
points. 

From the analyses of  Figures 7 and 8 , one can deduce that  
although the heavy scalar particles have definite CP parities  for
small $\tan\beta$ values,
they do not have definite CP characteristics  for  large $\tan\beta$  and  
differ from the lightest Higgs boson  in a sense that not
only they have   different masses  but also they have undefinite CP characteristics.

\begin{figure}[ht]
\centerline{\epsfig{file=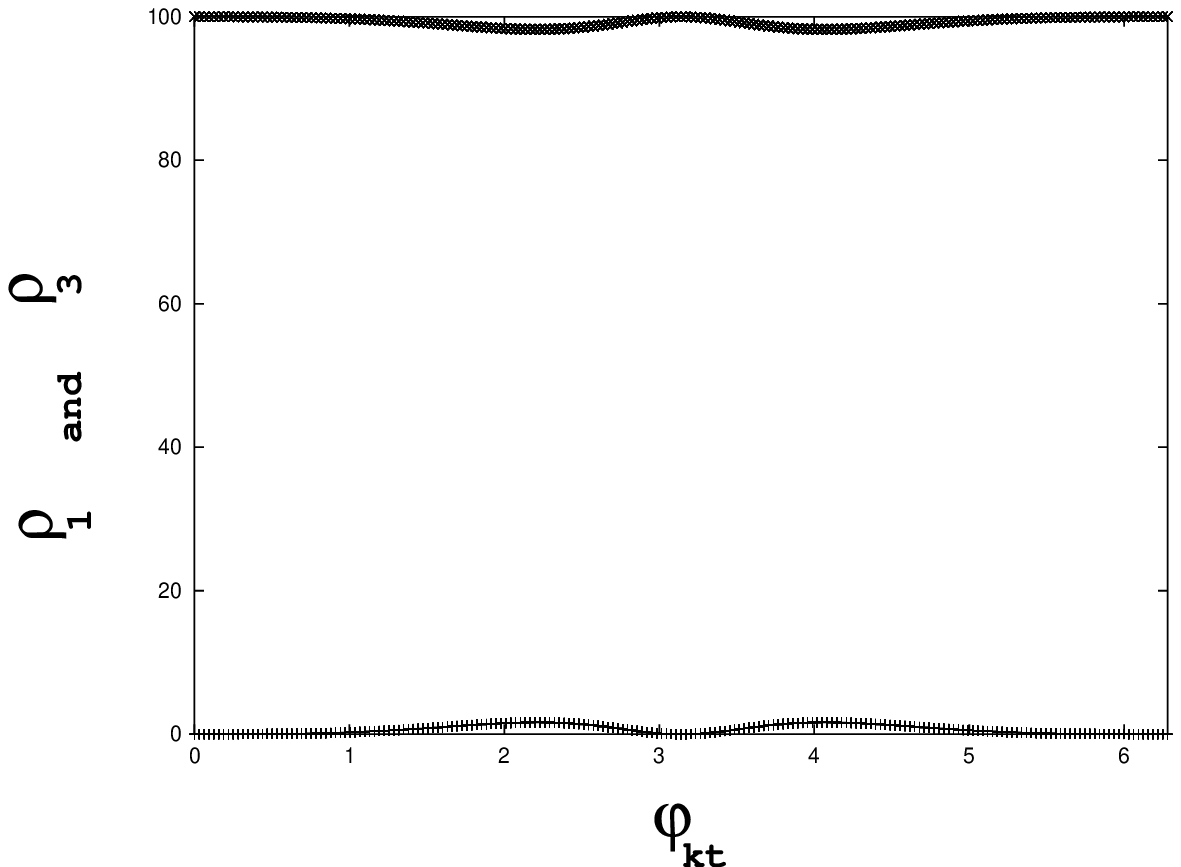,height=2.6in,width=2.6in }
\epsfig{file=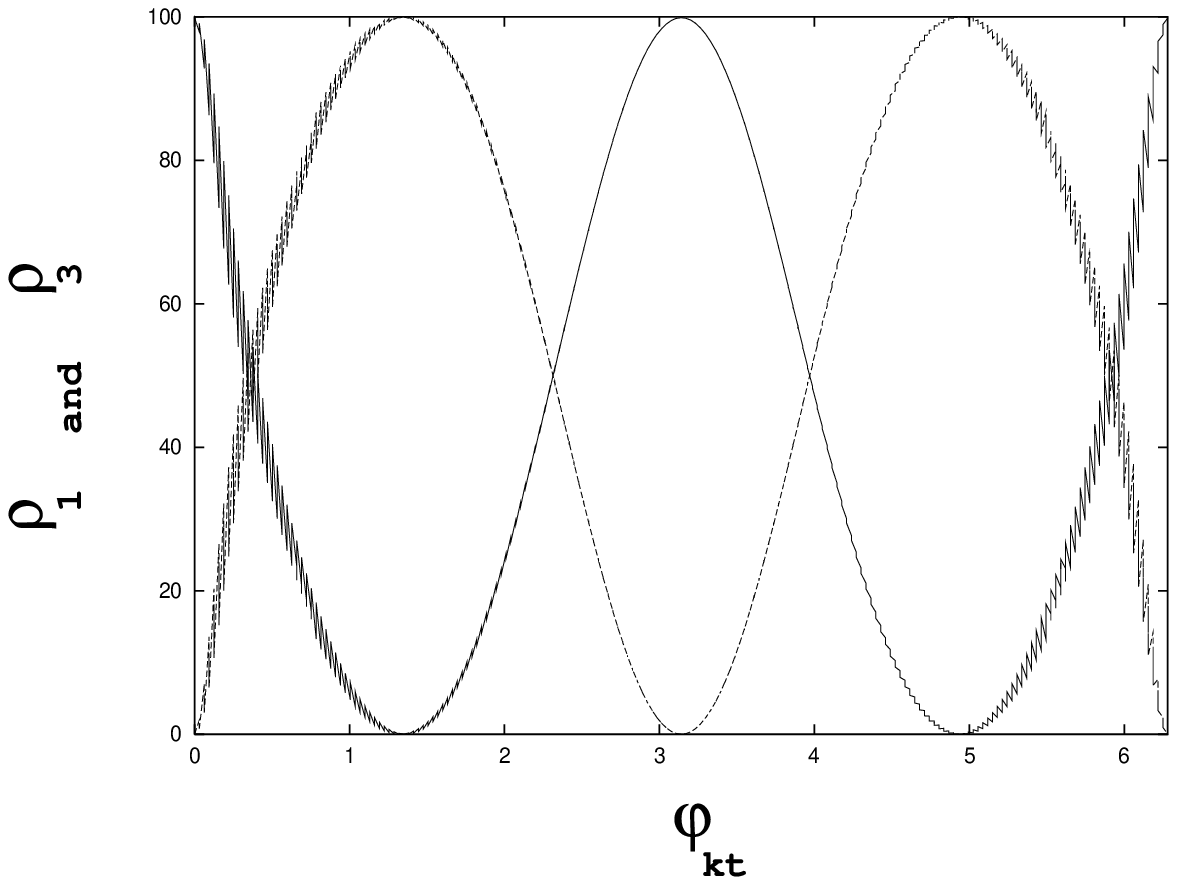,height=2.6in,width=2.6in }}
\caption{  $\rho_{1}$ (bottom curve ) and $\rho_{3}$ (top curve ) components of
$h_{2}$ as a function of $\varphi_{kt}$ 
for  $\tan\beta=4$  ( left panel) and  $\tan\beta=30$ ( right
panel).} 
\end{figure}
\begin{figure}[ht]
\centerline{\epsfig{file=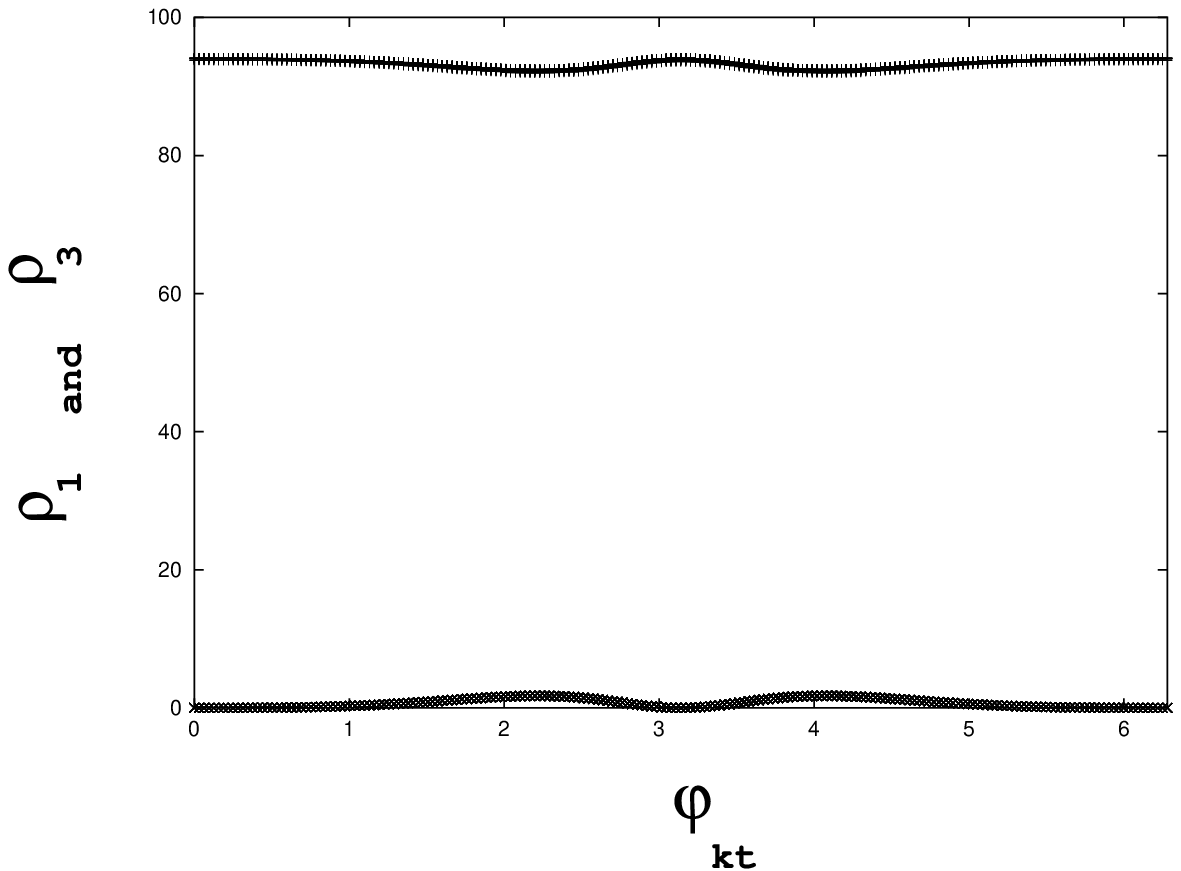,height=2.6in,width=2.6in }
\epsfig{file=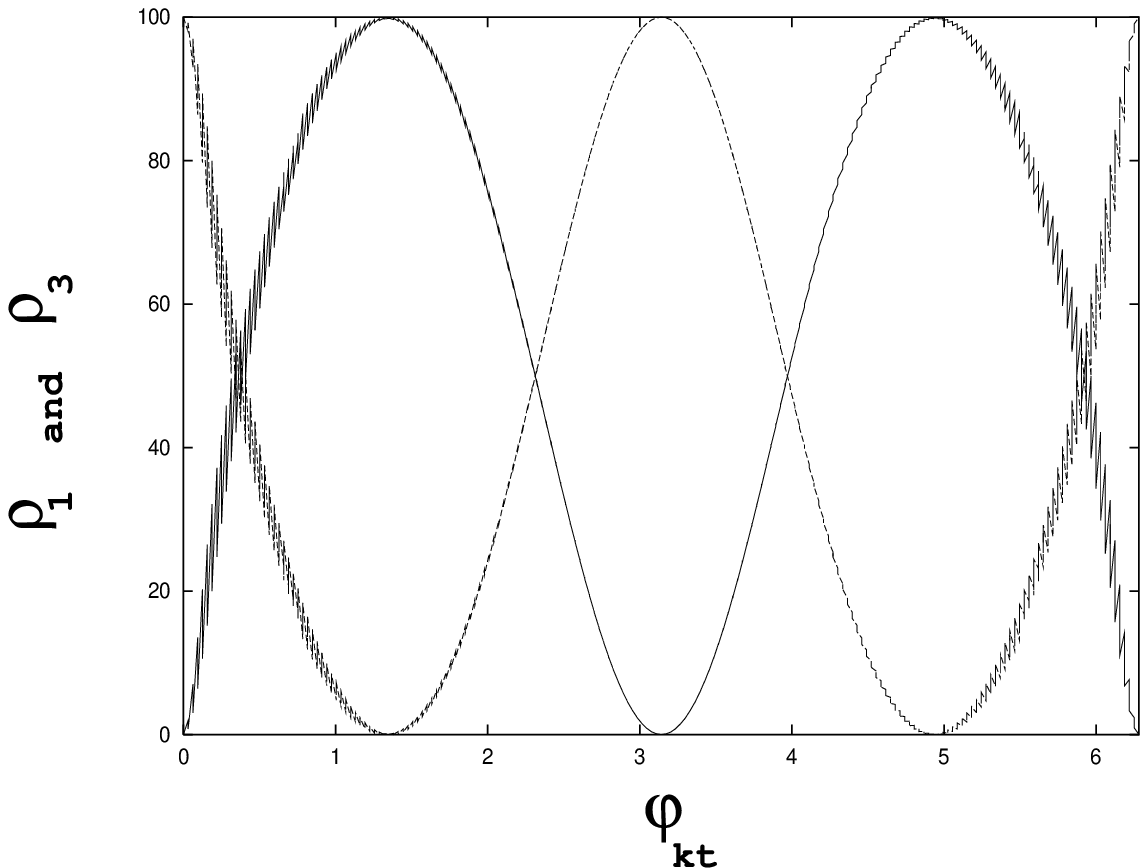,height=2.6in,width=2.6in }}
\caption{  $\rho_{1}$ (top curve ) and $\rho_{3}$ (bottom curve ) components of
$h_{1}$ as a function of $\varphi_{kt}$ 
for  $\tan\beta=4$  ( left panel) and  $\tan\beta=30$ (right
panel).} 
\end{figure}

\section{Conclusion}
From the analyses of the masses and the CP compositions of the  
Higgs bosons, we conclude that: 

$(i)$ The lightest Higgs mass is quite sensitive to the SUSY CP phases,
thanks to which the there arise new regions of the SUSY parameter space in which the present experimental
constaints are satisfied,

$(ii)$
Although the percentage CP-odd composition  of the lightest Higgs 
increases relatively with the increasing 
$\tan\beta$, it is still very small as compared to its CP--even compositions.
The lower limit on the lightest Higgs mass is directly correlated with
its CP--odd composition.

$(iii)$
The percentage CP-odd composition  of the lightest Higgs (with a mass
$m_{h_{3}}\simgt 115~{GeV}$) can not gain an appreciable value
unless $\mu$ is chosen smaller. For smaller values of $\mu$, 
however, the parameter space is constrained by the
existing LEP bound on the lightest Higgs \cite{16}. 

$(iii)$ The remaining two heavy scalars have definite CP-parities in the
small $\tan\beta$ regime, but they have no definite CP-parities in the large
$\tan\beta$ regime.
In this sense, they differ from the lightest Higgs as to
their masses and their undefinite  CP characteristics.

$(iv)$ The gluino axion model, besides solving the strong CP and 
$\mu$--problems in an economical way, provides a quite restricted parameter 
space due to the naturalness requirements.

\section{Acknowledgements}
This work was partially supported by the  Scientific and
Technical Research Council of Turkey (T\"{U}B{\.I}TAK) under the project,
No:TBAG-2002(100T108).

\end{document}